\documentclass[preprint,aps,showpacs,showkeys,tightenlines,superscriptaddress]{revtex4}
\usepackage{epsfig}
\newcommand{\bfi}[1]{\mbox{\boldmath $#1$}}
\newcommand{\bfis}[1]{\mbox{\boldmath ${\scriptstyle #1}$}}
\begin{document}
\title{Structure of and $E$2 transition in $^{16}$C 
in a $^{14}$C+$n$+$n$ model}
\author{W. Horiuchi}
\affiliation{Graduate School of Science and Technology, Niigata
University, Niigata 950-2181, Japan}
\author{Y. Suzuki}%
\affiliation{Department of Physics, Niigata University, Niigata
950-2181, Japan}
\pacs{23.20.-g, 21.10.-k, 21.60.-n, 27.20.+n}
\keywords{$^{16}$C; $E$2 transition; three-body model}

\begin{abstract}
A three-body model of $^{14}$C+n+n is applied to study the energy 
spectrum and the hindered $E$2 transition in $^{16}$C. A realistic 
two-nucleon potential is used for the valence neutrons. Both spin 
singlet and triplet components for the neutrons are taken into 
account. The three-body problem with a Pauli constraint is solved 
in a stochastic variational method. 
For the n-$^{14}$C potential chosen to reproduce the properties of 
$^{15}$C, the low-lying energy spectrum agrees reasonably well 
with experiment, but the ground state is predicted to be about 1 MeV 
high. The calculated $B(E2$;\, $2_1^+ \!\to \!0^+_1)$ value is 
about twice the measured value if the polarization charge of 
the valence neutrons is taken to be the same as that required 
to fit the $^{15}$C data.  The correlated motion of the valence 
neutrons is displayed through the two-neutron density distribution.

\end{abstract}
\maketitle
\draft

Recently the $E2$ transition 
from the first 2$^+$ state to the ground 0$^+$ state 
in $^{16}$C 
has been studied through a lifetime measurement using 
a recoil shadow method~\cite{imai} and $^{16}$C+$^{208}$Pb 
inelastic scattering~\cite{elekes}. 
The $B(E2)$ value is found to be $0.63\pm 0.12$ $e^2\, $fm$^4$, 
which corresponds to anomalously small strength of about 
0.26 W.u. 
The anomaly is apparent by comparison with other C isotopes. 
The $B(E2$;\, 2$_1^+\to 0_1^+$) values of $^{10}$C, $^{12}$C 
and $^{14}$C decrease gradually as expected from the increasing 
excitation energies. However, the case of 
$^{16}$C strongly deviates from this expectation: 
The excitation energy of the 2$^+_1$ state of $^{16}$C is only 
1.77 MeV, much smaller than the case of $^{14}$C (7.01 MeV), but 
the $B(E2)$ value of $^{16}$C is nevertheless much 
smaller than that of $^{14}$C, which is 3.74 $e^2$fm$^4$ 
(1.87 W.u.). The $B(E2)$ value for the $\frac{5}{2}^+_1\to
\frac{1}{2}^+_1$ transition in $^{15}$C is also small, 
$0.97\pm 0.02$ $e^2\, $fm$^4$ (0.44 W.u.)~\cite{selove}.

In the previous paper~\cite{suzukic16}, a three-body 
model of $^{14}$C+n+n was applied to study the anomalous $E$2 
transition in $^{16}$C under the assumption that the $^{14}$C core has a 
neutron shell closure of the p shell and the two valence neutrons are in 
a spin-singlet state. 
The $^{14}$C+(sd)$^2$ model for the low-lying levels of $^{16}$C is
supported by experiment as well~\cite{fortune2,bohlen}. 
The calculated $B(E2)$ value was 1.38 $e^2\, $fm$^4$, 2.2 times 
larger than experiment  
if one uses the same neutron polarization charge as that required 
for $^{15}$C. The deformations of both proton and neutron densities 
of the C isotopes are studied in antisymmetrized molecular dynamics 
calculation~\cite{kanada}, where the neutron shape is found to 
change significantly with the neutron number and the $B(E2)$ value 
of $^{16}$C in a variation after projection calculation is 
4-6 times too large compared to experiment. 
The $B(E2)$ value is overestimated also in the deformed Hartree-Fock 
model and the shell model calculations~\cite{sagawa}.   
The purpose of the present investigation is to study the structure 
of $^{16}$C thoroughly and to reexamine a mechanism which leads 
to the hindered transition. Our basic assumption is the same as before, so 
the relevant levels of $^{16}$C are assumed to be generated 
from the $^{14}$C+n+n model. There are, however, some noticeable 
differences. We here use more general basis 
functions allowing for both spin-singlet and spin-triplet 
neutrons. This makes it possible to obtain unnatural parity 
states such as 3$^+$. In addition, we use a realistic 
interaction for the two neutrons in order to avoid uncertainties 
of the model, while in the previous paper an effective 
interaction was used and its strength was increased so as to 
fit the ground-state energy.

The wave function for $^{16}$C is determined 
from the following Hamiltonian 
\begin{equation}
H=T_{\bfis R}+T_{\bfis r}+U_1+U_2+v_{12},
\end{equation}
where the subscripts of the kinetic energies stand for 
the relative distance vector, ${\bfi R}$, from the 
center-of-mass of $^{14}$C 
to that of the two neutrons and the relative distance
vector of the neutrons, ${\bfi r}$. The two-neutron potential $v_{12}$ 
is taken from G3RS (case 1) potential~\cite{tamagaki} which 
contains central, tensor and spin-orbit forces 
and reproduces the nucleon-nucleon scattering data 
as well as the deuteron properties. 
The n-$^{14}$C potential $U$ takes the form 
\begin{equation}
U=-V_0f(r)+V_1{\bfi \ell}\cdot{\bfi s}{\frac{1}{r}}{\frac{d}{dr}}f(r),
\label{opt.pot}
\end{equation}
where $f(r)$=$[1+{\rm exp}({\frac{r-R_{\rm c}} {a}})]^{-1}$ 
with $R_{\rm c}$=$r_0A_{\rm c}^{\,\frac{1}{3}}\, (A_{\rm c}\!=\!14)$. 
The parameters of $U$ are determined to reproduce 
the energies of the $\frac{1}{2}^+_1$ and $\frac{5}{2}^+_1$ states 
of $^{15}$C. 
In order to make the $1s_{{1}/{2}}$ state lower than the 
$0d_{{5}/{2}}$ state, the ${\bfi \ell}\cdot{\bfi s}$ strength $V_1$ 
was chosen in the previous study 
to be about half of the standard value~\cite{bm}. 
This potential is called 
set A hereafter. We also test other possibilities to fit these  
single-particle (s.p.) energies. In set B we use the standard 
value of $V_1$ but 
weaken $V_0$ for all the partial waves other than $s$-wave, 
while in set C a potential with a larger diffuseness parameter 
is chosen. 
The parameters of each set are listed in Table~\ref{c14-n.pot}. 
These potentials generate almost the same s.p. wave function for 
both $1s_{{1}/{2}}$ and $0d_{{5}/{2}}$. 

\begin{table}[t]
\caption{Parameters of n-$^{14}$C potential.}
\label{c14-n.pot}
\begin{center}
\begin{tabular}{cc|ccccccc}
\hline\hline
&  &&$V_0$(MeV) & $V_1$(MeV\,fm$^2$) & & $r_0$(fm) & & $a$(fm)\\
\hline
set A && &$-$50.31            & 16.64 & & 1.25 & & 0.65 \\
set B && &$-$50.31\,($l\!=\!0$),\,$-$47.18\,($l\!\neq\!0$)& 31.25 &&1.25 &&0.65\\
set C && & $-$51.71            & 26.24 & &1.20 & &0.73 \\
\hline\hline
\end{tabular}
\end{center}
\end{table}

Trial wave functions are 
assumed as a combination of correlated Gaussian bases 
\begin{eqnarray}
& &\qquad \Psi_{JM}(1,2)=\sum_{i=1}^KC_i \Psi_{JM}(\lambda_i,A_i),\\
& &\Psi_{JM}(\lambda,A)=(1-P_{12}) 
 \left\{ {\rm e}^{-{\frac{1}{2}}
\tilde{\bfis x}A{\bfis x}} [[{\cal Y}_{\ell_1}({\bfi x}_1)
{\cal Y}_{\ell_2}({\bfi x}_2)]_L\chi_S(1,2)]_{JM}\right\},
\label{base}
\end{eqnarray}
where $P_{12}$ is a permutation of the 
two neutrons. The basis function is specified by a set of angular 
momenta $\lambda\!=\!(\ell_1,\ell_2,L,S)$ and a 2$\times$2 
positive-definite, symmetric 
matrix $A$. Here $\tilde{\bfi x}A{\bfi x}$ stands for 
$A_{11}{\bfi x}_1^{\, 2}$+$2A_{12}{\bfi x}_1
\!\cdot\!{\bfi x}_2$+$A_{22}{\bfi x}_2^{\, 2}$, and 
${\bfi x}_1$=${\bfi R}$+${\frac{1}{2}}{\bfi r}$ and ${\bfi x}_2$
=${\bfi R}$$-$${\frac{1}{2}}{\bfi r}$ 
are the distance vectors of the neutrons from the 
center-of-mass of the $^{14}$C core. The two neutrons 
are explicitly correlated due to the presence of the cross term 
$A_{12}{\bfi x}_1$$\cdot$${\bfi x}_2$, the inclusion of which 
is very important to obtain a precise 
solution~\cite{svm,book}. 
The angular parts of the two-neutron orbital motion are 
described using 
${\cal Y}_{\ell m}({\bfi r})\!=\!r^{\ell}Y_{\ell m}(\hat{\bfi r})$ and 
they are coupled 
with the spin part $\chi_S$ to the total angular momentum $J$. 
In the previous study, the angular motion was described 
with much simpler functions specified by a global 
vector~\cite{global}, but here a general form is used to take into
account important $L,S$ contents of the wave functions. 

The energy and the corresponding wave function are determined 
from a solution of the generalized eigenvalue problem 
\begin{eqnarray}
& &\sum_{j=1}^K [H_{ij}-EB_{ij}]C_j=0\ \ \ \ \ (i=1,2,\ldots,K),
\\
& &\left( \!
\begin{array}{c}
H_{ij} \\
B_{ij} 
\end{array} \! \right)
=\langle \Psi_{JM}(\lambda_i, A_i)\mid \!
\left( \!
\begin{array}{c}
H \\
1 
\end{array} \!\right) \!
\mid\Psi_{JM}(\lambda_j, A_j)\rangle.
\end{eqnarray}
It is vital to take into account the Pauli principle for the 
motion of the valence neutrons. Under the assumption that 
$^{14}$C is neutron-closed, the Pauli constraint is fulfilled 
by imposing that the trial wave function has no overlap with 
any orbit $u_{n\ell jm}$ occupied in the $^{14}$C core
\begin{equation}
\langle u_{n\ell jm}(i)|\Psi_{JM}(1,2) \rangle=0\ \ \ \ \ (i=1,2),
\label{pauli}
\end{equation} 
where the s.p. orbit $u_{n\ell jm}$ is 
generated from $U$ and $n\ell j$ runs over 
$0s_{{1}/{2}},\, 0p_{{3}/{2}}$, and $0p_{{1}/{2}}$. 
The condition~(\ref{pauli}) is practically achieved 
by the orthogonal projection method~\cite{kukulin}. 
The accuracy of the present calculation is such that 
the probability of mixing-in of the 
occupied orbits is of the order of 10$^{-4}$. 

Before superposing basis functions with different $\lambda$, 
we did a pilot calculation in a single $\lambda$ channel. 
We use the algorithm called the stochastic 
variational method (SVM)~\cite{svm,book} to optimize the 
parameter matrices $A$. The SVM increases the basis dimension 
one by one by testing a number of candidates which are chosen 
randomly, and in addition fine-tunes the already chosen parameters 
by a refinement process. The basis selection with the SVM is 
effectively performed not only to take care of the short-range 
repulsion of $v_{12}$ but also to satisfy 
the condition~(\ref{pauli}). For $J^{\pi}$=0$^+$, those channels 
which give an energy lower than the $^{14}$C+n+n threshold are only
$\lambda$=$(0,0,0,0)$ and $(2,2,0,0)$, while for $J^{\pi}$=2$^+$ 
such a bound solution is obtained only for $\lambda$=$(0,2,2,0)$. 
Channels with $\ell_i \ge 3$ play a minor role. 
We also studied levels 
with $J^{\pi}$=3$^+$ and 4$^+$. The most important channel is 
$\lambda$=$(0,2,2,1)$ for 3$^+$ and $\lambda$=$(2,2,4,0)$ and 
$(2,2,3,1)$ for 4$^+$, respectively. Thus the $S\!=\!1$ 
channel is important for 4$^+$.

\begin{figure}[b]
\epsfig{file=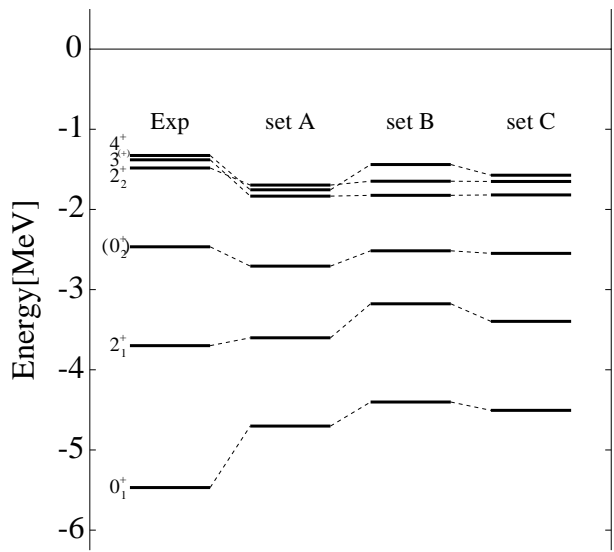,width=8.8cm,height=7.2cm}
\caption{Low-lying energy levels of $^{16}$C for different sets 
of the n-$^{14}$C potential. Energy is 
given from the $^{14}$C+n+n threshold.}
\label{eng.fig}
\end{figure}

The basis sets chosen in the single-channel calculation serve as  
the basis functions for a coupled-channels calculation. 
Channels are truncated to those 
with $\ell_1\!+\!\ell_2 \!\le$4. A basis dimension $K$ used in the 
calculation is about 450 for 0$^+$ and 500-650 for the other 
cases. Figure~\ref{eng.fig} compares the calculated energy levels 
of $^{16}$C with experiment. The $U$-dependence of the 
spectrum is moderate. The ground-state energy 
is about 1 MeV too high compared to experiment, but 
the energies of the other 
$J^{\pi}$ states are in fair agreement with experiment. 
No state with $J^{\pi}$=1$^+$ was obtained below the 
$^{15}$C+n threshold. 

Table~\ref{c15.16} summarizes some properties of the 
states calculated with set B potential. The other potentials give 
similar results. The root mean square radius of 
$^{16}{\rm C}$, $r(^{16}{\rm C})$, for the point nucleon 
distribution is calculated from the equation, 
$r^2(^{16}{\rm C})$=${\frac{14}{16}}r^2(^{14}{\rm C})$+${\frac{7}{64}}
\langle {\bfi R}^2 \rangle$+${\frac{8}{256}}
\langle {\bfi r}^2 \rangle$, where we substitute 
the point proton radius, 2.35 fm~\cite{selove} for 
$r(^{14}{\rm C})$. The $r(^{16}{\rm C})$ value for the ground 
state agrees well with that extracted from the reaction 
cross section analysis~\cite{zheng}. 
The probability of spin-singlet components, $P_{S=0}$, 
shows that our previous calculation~\cite{suzukic16} 
with no $S\!=\!1$ components is not very good 
for the $2^+_1$ state but quite acceptable for the ground state.  
$P_{ss},P_{sd}$ and $P_{dd}$ stand for the 
probabilities for the two neutrons to occupy the 
$(1s_{{1}/{2}})^2, \, (1s_{{1}/{2}}0d_{{5}/{2}})$ and 
$(0d_{{5}/{2}})^2$ configurations, respectively. 
The sum of the probabilities, $P_{ss}\!+\!P_{sd}\!+\!P_{dd}$, 
is less than unity, which signals the importance of unbound 
s.p. orbits or n+$^{14}$C continuum states. Its effect appears 
more important in the $0_1^+$ and $2_1^+$ states. 
The previous calculation gave $P_{ss}\!=\!0.49$ and 
$P_{dd}\!=\!0.39$ for the ground state, which is consistent with 
the present result. We thus expect that 
the momentum distribution data of 
$^{15}$C fragments from $^{16}$C breakup~\cite{yamaguchi} are 
well reproduced in the present model 
as in the previous paper~\cite{suzukic16}.

\begin{table}[t]
\caption{Properties of the states in $^{16}$C. Set B parameters 
are used. Energy and length are in MeV and fm, respectively.}
\label{c15.16}
\begin{center}
\begin{tabular}{cccccccccccccccc}
\hline\hline
$J^{\pi}$ &   & $E_{\rm cal}$ &   & $E_{\rm exp}$  
& & $\langle{\bfi x}_1^2\rangle$  &   & $\langle {\bfi R}^2\rangle$
& $\langle {\bfi r}^2\rangle$
& $\langle {\bfi x}_1$$\cdot$ $ {\bfi x}_2\rangle$ & $r$($^{16}$C) 
& $P_{S=0}$ & $P_{ss}$ & $P_{sd}$ & $P_{dd}$\\
\hline
0$^+_1$  &    & $-$4.40  &   & $-$5.469 & & 17.8  &  & 10.4  & 29.5  & 3.04  & 2.63 & 0.92 & 0.52 & -- & 0.37\\
0$^+_2$  &    & $-$2.52  &    & $-$2.466 & &19.7  &  & 10.4  & 37.3   &
 1.10  & 2.67 & 0.86 & 0.43 & -- & 0.53 \\
2$^+_1$ &    & $-$3.18 &    & $-$3.699 & &16.8  &   & 9.39  & 29.7  
&  1.95  &  2.61  &  0.75 & --& 0.67 & 0.24 \\
2$^+_2$ &    & $-$1.65 &    & $-$1.483 & &17.1  &   & 8.95  & 32.5  &  0.84  &  2.62  &  0.55 & -- & 0.26 & 0.72\\
3$^+_1$ &    & $-$1.82 &    & $-$1.381 & &22.3  &   &11.5  & 43.1  
&  0.73  &  2.73  &  0.00 & -- & 0.99 & --\\
4$^+_1$ &    & $-$1.44 &    & $-$1.327 & &15.9  &   & 8.30  & 30.4  &  0.70  &  2.59  &  0.35 & -- & -- & 0.96 \\
\hline\hline
\end{tabular}
\end{center}
\end{table}

\begin{figure}[b]
\epsfig{file=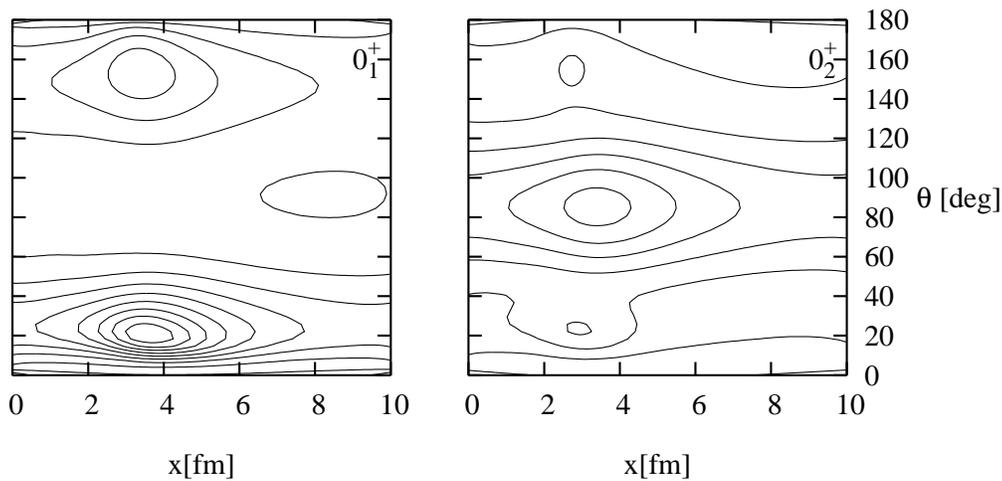,width=16.0cm,height=6.7cm}
\caption{Contour maps of the two-neutron densities 
$\rho(x,x,\theta)$, 
weighted by $8\pi^2x^4{\rm sin}\theta$, for the $0^+_1$ 
and $0^+_2$ states of $^{16}$C. The contour shows peaks, 
and the difference between any two neighboring contour levels is 0.002 
fm$^{-2}$. See text.}
\label{density}
\end{figure}

The two-neutron correlation is examined with the 
density distribution function
\begin{equation}
\rho(x_1,x_2,\theta)=\frac{1}{2J+1}\sum_M\langle \Psi_{JM}(1,2)\mid
 \Psi_{JM}(1,2) \rangle_{\rm spin},
\end{equation}
where $\theta$ is the angle between ${\bfi x}_1$ and ${\bfi x}_2$ 
and $\langle\cdots \rangle_{\rm spin}$ indicates that the 
integration is to be done over the spin coordinates only. 
Figure~\ref{density} displays the contour map of 
$8\pi^2x^4{\rm sin}\theta\,\rho(x,x,\theta)$ for the $0^+$ 
states. For the $0^+_1$ state, we have two distinct peaks: 
One peak with smaller angles is the highest (about 0.02 fm$^{-2}$ in 
height), suggesting the 
correlation of ``di-neutron'' type, and the other with larger 
angles reaches half the highest peak, 
corresponding to a ``cigar''-like configuration in which the two neutrons 
sit on the opposite sides of the core. For the $0^+_2$ state, 
however, only one distinct peak with a height of about 
0.014 fm$^{-2}$ appears around $\theta\!=\!90^{\circ}$ 
(a ``boomerang'' shape). These 
different characteristics are expected from the expectation 
value of ${\rm cos}\, \theta$, which is estimated from 
$\langle {\bfi x}_1\!\cdot \!
{\bfi x}_2\rangle$/$\langle{\bfi x}_1^2\rangle$. This value 
is 0.17 for $0^+_1$ and 0.06 for $0^+_2$, as seen from 
Table~\ref{c15.16}. We also confirmed 
that the $2^+_1$ state has two peaks similarly  
to the ground state but the other states all have one peak 
at the boomerang configuration.

\begin{figure}[b]
\epsfig{file=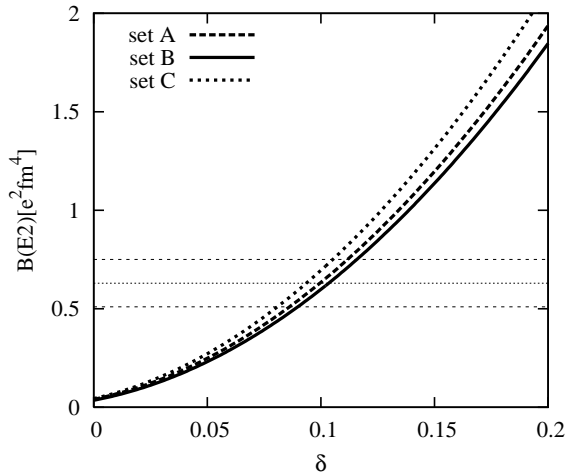,width=7.8cm,height=6.5cm}
\caption{The $B(E2)$ value for the transition from the 2$_1^+$ 
state to the ground state 
in $^{16}$C as a function of the polarization 
charge $\delta$. The measured value~\protect\cite{imai,elekes}
$0.63\pm 0.12$ $e^2\, $fm$^4$ is indicated by horizontal lines.  }
\label{be2.fig}
\end{figure}

The $E2$ operator, 
${\cal M}_{\mu}$, for $^{16}$C in the $^{14}$C+n+n
model is expressed as~\cite{suzukic16}
\begin{equation} 
{\cal M}_{\mu}=\frac{1}{2}\,\delta\,e\,{\cal Y}_{2\mu}({\bfi r})+q_{\rm eff}\,
e\, {\cal Y}_{2\mu}({\bfi R}),
\label{e2op}
\end{equation}
with $q_{\rm eff}$=$\frac{3}{32}$+$\frac{49}{32}\delta$, where  
$\delta$ is a polarization charge of the valence neutron. 
The value of $\delta$ can be estimated so as to 
fit the $B(E2)$ value of $^{15}$C within the $^{14}$C+n model, 
which leads to $\delta$=0.160 (set A), 0.159 (set B) and 0.151 
(set C), respectively. 
The first term on the right-hand side of Eq.~(\ref{e2op}) is the 
$E2$ operator corresponding to the relative motion of the two 
neutrons, while the second term corresponds to the relative 
motion between the $^{14}$C core and the center-of-mass of the 
two neutrons. If the valence neutrons have no 
charge ($\delta\!=\!0$), the $E2$ operator reduces to the 
second term of Eq.~(\ref{e2op}) with a small $q_{\rm eff}$. 
In this case, the $B(E2)$ value is contributed by the difference 
between the center of mass of $^{14}$C and that of $^{16}$C. For 
nonzero $\delta$, both terms of Eq.~(\ref{e2op}) contribute to 
the $E2$ transition matrix element. 
The $B(E2; 2^+_1\! \to \!0^+_1)$ value calculated with the 
different potentials are shown in Fig.~\ref{be2.fig}. 
All the cases give similar results. The $B(E2)$ value is 
about twice the measurement for such $\delta$ that is 
determined from the $^{15}$C data. It is interesting to see 
how the $B(E2)$ value changes with the admixture of the 
$S\!=\!1$ components. The $E2$ operator (\ref{e2op}) is 
independent of spin, so its reduced matrix element consists of two 
parts:
\begin{equation}
\langle 0^+_1||{\cal M}||2^+_1\rangle=\sqrt{(1-a^2)(1-b^2)}
\, \langle{\cal M}(S=0)\rangle+
ab\, \langle {\cal M}(S=1)\rangle,
\end{equation}
where $a$ and $b$ are the probability amplitudes of the $S\!=\!1$ 
components in the $0^+_1$ and $2^+_1$ states, respectively, and 
the $\langle{\cal M}(S)\rangle$ stands for the reduced matrix element 
between their (normalized) wave function components with spin $S$. 
For set B and for $\delta\!=\!0.159$, 
we have $\langle{\cal M}(S\!=\!0)\rangle\!=\!2.75\, e\,$fm$^2$ and 
$\langle{\cal M}(S\!=\!1)\rangle\!=\!1.56\, e\,$fm$^2$. If both of the 
$0^+_1$ and $2^+_1$ states are purely spin-singlet ($a\!=\!b\!=\!0$), 
the $B(E2)$ value 
would be $2.75^2/5\!=\!1.51 e^2\, $fm$^4$. In fact, they have $S\!=\!1$ 
admixtures whose 
magnitudes are $a^2\!=\!0.08$ and $b^2\!=\!0.25$ (see
Table~\ref{c15.16}), and therefore the $B(E2)$ value reduces to 
1.25 $e^2\, $fm$^4$, as shown in Fig.~\ref{be2.fig}. 
In the previous calculation with $S\!=\!0$ only~\cite{suzukic16}, 
$\langle {\cal M}(S=0)\rangle$ was 2.63 $e\, $fm$^2$, leading to the
$B(E2)$ value of 1.38 $e^2\, $fm$^4$. 
Thus the $S\!=\!1$ admixture has led to reducing 
the $B(E2)$ value.  

We have studied the structure and anomalous $E2$ transition 
of $^{16}$C in the $^{14}$C+n+n model, where the two valence 
neutrons interact via the realistic potential. We have tested 
three different n-$^{14}$C potentials which all reproduce the 
energies of the $\frac{1}{2}^+_1$ and $\frac{5}{2}^+_1$ states 
of $^{15}$C in the $^{14}$C+n model. Both spin singlet 
and triplet components are included in the calculation. 
The stochastic variational method has been used to solve the 
three-body problem with the Pauli constraint on the valence 
neutrons. 
The energy spectrum of $^{16}$C below the $^{14}$C+n+n threshold 
is in fair agreement with experiment except 
that the ground-state energy is about 1 MeV too high. 
The dependence of the $B(E2$;\,$2_1^+\! \to\! 0^+_1)$ value 
on the polarization charge of the valence neutrons is studied 
and found not to be very 
sensitive to the choice of the n-$^{14}$C potential. 
We have seen that admixing the $S\!=\!1$ components in the wave 
functions reduces the $B(E2)$ value considerably. The 
calculated $B(E2)$ value is, however, about twice the observed value 
if the polarization charge is set to reproduce the 
$B(E2$;\,$\frac{5}{2}^+_1\! \to\! \frac{1}{2}^+_1)$ of $^{15}$C. 

Within the $^{14}$C+n+n model, the present calculation is probably 
the most unambiguous because it is free from any adjustable parameters and 
the result is rather insensitive to the choice of the n-$^{14}$C
potential. The fact that the ground-state energy 
is predicted to be 1 MeV high and that the theoretical $B(E2)$ value is 
still twice large may point to the necessity of other
mechanisms such as core distortion or excitation. 

\vspace{5mm}

The authors thank H. Matsumura and B. Abu-Ibrahim for 
communication. This work is in part supported by JSPS 
Grants-in-Aid for Scientific Research 
(No. 14540249) and a Grant for Promotion of Niigata University 
Research Projects (2005-2007).

\end{document}